\documentstyle[prc,aps]{revtex}
\begin{document}
\draft
\title{ Effective nucleon mass in relativistic mean field theory}
\author{K.C. Chung$^1$, C.S. Wang$^{1,2}$, A.J. Santiago$^1$, and 
J.W. Zhang$^2$}

\address{\it (1) Instituto de F\'\i sica, Universidade do Estado do Rio 
de Janeiro,\\
Rio de Janeiro-RJ 20559-900, Brazil \\
(2) Department of Technical Physics, Peking University,
Beijing 100871, China}
\date{\today}
\maketitle
\begin{abstract}
In the $\sigma$-$\omega$-$\rho$ model of the relativistic mean field 
theory  with nonlinear $\sigma$-meson self-interaction, the effective 
nucleon mass $M^*$ is discussed with relation to the symmetry 
incompressibility $K_s$ of nuclear matter, based on the model parameters 
fitted to nuclear matter properties. It is shown that $M^*$ is larger 
than $0.73M$, if $K_s$ is assumed to be negative and the nuclear matter 
incompressibility $K_0$ is kept less than $300 MeV$. Furthermore, the 
field system is shown to be stable, as the $\sigma$-meson 
self-interaction energy is lower bounded in this parameter region.
\end{abstract}


As starting point for the relativistic microscopic description of the 
nuclear many-body system, within the framework of quantum hadrodynamics, 
the well-studied $\sigma$-$\omega$-$\rho$ model with nonlinear 
$\sigma$-meson self-interaction has been proved to be able to describe 
the saturation and other properties of nuclear matter\cite{Serot97}. 
However, there is some controversy about the stability of the field 
system specified by the fitted parameters, as the fitted coefficient of 
the fourth power term of the $\sigma$-meson field in the Lagrangian 
density is negative, and thus the energy of the field system diverges 
for large $\sigma$-field\cite{Serot97}\cite{Reinhard89}. In addition, 
the symmetry incompressibility $K_s$ of nuclear matter determined by the 
parameter sets existing in the market is positive, which is opposite to 
that given by the nonrelativistic models of nuclei\cite{CWSZ}. 
Furthermore, the effective nucleon mass $M^*$ is around $0.6M$ which 
seems uncomfortably low. Then, the relevant question is: are these 
features manifestation of intrinsic properties of the model itself, or 
do they depend on the input data used for fitting the model parameters? 
The purpose of this rapid communication is to show that the stable 
result with negative symmetry incompressibility $K_s$ and larger 
effective nucleon mass $M^*$ can be obtained, if the model parameters 
are fitted to reasonable nuclear matter properties.
 
The $\sigma$-$\omega$-$\rho$ model of the relativistic mean field theory 
is specified by the following Lagrangian density\cite{Serot97}(we use 
natural units with $\hbar=c=1$):
$${\cal L}=\overline\psi[\gamma_\mu(i\partial^\mu
-g_\omega\omega^\mu-g_\rho\mbox{\boldmath $\tau\cdot$}{\bf b}^\mu)
-(M-g_\sigma\phi)]\psi$$
$$+\frac 12(\partial_\mu\phi\partial^\mu\phi
-m_\sigma^2\phi^2)-\frac 13Mb(g_\sigma\phi)^3
-\frac 14c(g_\sigma\phi)^4$$
$$-\frac 14F_{\mu\nu}F^{\mu\nu}+\frac 12m_\omega^2\omega_\mu\omega^\mu$$
\begin{equation}\label{sorL}
-\frac 14{\bf B}_{\mu\nu}\mbox{\boldmath $\cdot$} 
{\bf B}^{\mu\nu}+\frac 12m_\rho^2{\bf b}_\mu\mbox{\boldmath $\cdot$}
{\bf b}^\mu,
\end{equation}
where $F^{\mu\nu}=\partial^\mu\omega^\nu-\partial^\nu\omega^\mu,$
${\bf B}^{\mu\nu}=\partial^\mu{\bf b}^\nu-\partial^\nu{\bf b}^\mu$; 
$\psi$, $\phi$, $\omega$ and ${\bf b}^\mu$ are the nucleon, $\sigma$, 
$\omega$ and $\rho$ meson fields with masses $M$, $m_\sigma$, $m_\omega$ 
and $m_\rho$, respectively, while $g_\sigma$, $g_\omega$ and $g_\rho$ 
are the respective meson-nucleon coupling constants; $b$ and $c$ are the 
nonlinear term coefficients, and {\boldmath $\tau$} are isospin 
matrices. As $M$, $m_\omega$ and $m_\rho$ are taken from experiment, the 
model parameters are $g_\sigma$, $g_\omega$, $g_\rho$, $b$, $c$ and 
$m_\sigma$. 

The nuclear matter equation of state derived from this Lagrangian density 
can be expressed in terms of the nuclear energy density ${\cal E}$ as 
$e={\cal E}/\rho_N-M$, where $\rho_N$ is the baryonic density, and
\begin{equation}\label{Eksor}{\cal E}={\cal E}_k
+{\cal E}_\sigma+{\cal E}_\omega+{\cal E}_\rho,
\end{equation}
\begin{equation}\label{Ek}{\cal E}_k
=\frac{M^4\xi^4}{\pi^2}\sum_{i=p,n}F_1(k_i/\xi M),
\end{equation}
\begin{equation}\label{Esigma}{\cal E}_\sigma
=M^4\Big[\frac 1{2C_\sigma^2}(1-\xi)^2
+\frac 13b(1-\xi)^3+\frac 14c(1-\xi)^4\Big],
\end{equation}
\begin{equation}\label{Eomega}{\cal E}_\omega
=\frac{C_\omega^2\rho_N^2}{2M^2},
\end{equation}
\begin{equation}\label{Erho}{\cal E}_\rho
=\frac{C_\rho^2\rho_N^2}{2M^2}\delta^2,
\end{equation}
where $\delta=(\rho_p-\rho_n)/\rho_N$ is the nuclear matter asymmetry; 
$k_p$ and $k_n$ are the proton and neutron Fermi momenta respectively, 
\begin{equation}\label{xi}\xi=\frac{M^*}M
=1-\frac{g_\sigma}M\phi,
\end{equation}
\begin{equation}\label{Ci}C_i=g_i\frac M{m_i},\,\,\,\,
i=\sigma, \omega, \rho, 
\end{equation}
and the function $F_m(x)$ is defined as (see Ref.\cite{CWSZ} for details):
\begin{equation}
F_m(x)=\int_0^xdx\,x^{2m}\sqrt{1+x^2}.
\end{equation}
The reduced effective nucleon mass $\xi$ and thus the field $\phi$ is 
determined by
\begin{equation}\label{Eqxi}(1-\xi)+bC_\sigma^2(1-\xi)^2
+cC_\sigma^2(1-\xi)^3=\frac{C_\sigma^2}{M^3}\rho_s,
\end{equation}
where the scalar density $\rho_s$ can be expressed as
\begin{equation}\label{rhos}\rho_s=\frac{M^3\xi^3}{\pi^2}
\sum_{i=p,n}f_1(k_i/\xi M),
\end{equation}
and the function $f_m(x)$ is defined as (see Ref.\cite{CWSZ} for details) 
\begin{equation}
f_m(x)=\int_0^xdx\frac{x^{2m}}{\sqrt{1+x^2}}.
\end{equation}
The model parameters related to the nuclear equation of state are only 
$C_\sigma^2$, $C_\omega^2$, $C_\rho^2$, $b$ and $c$. As $m_\sigma$ is 
the inverse Compton wavelength of the $\sigma$-meson, it is related only 
to finite size effects, such as surface energy and shell effects of 
nuclei. 

Knowing the equation of state, the following formula for pressure $p$ 
can be obtained:
\begin{equation}\label{pressure}p=-{\cal E}
+\rho_N\frac{\partial{\cal E}}{\partial\rho_N}
=\frac 13{\cal E}_k-\frac 13M\xi\rho_s-{\cal E}_\sigma
+{\cal E}_\omega+{\cal E}_\rho.
\end{equation}
The standard state $(\rho_N=\rho_0, \delta=0)$ is defined by the 
equilibrium condition, $p\,(\rho_0, 0)=0$. The standard density $\rho_0$ 
can be written in terms of nuclear radius constant $r_0$ or nucleon 
Fermi momentum $k_F$ as
\begin{equation}\label{rho0r0kF}
\rho_0=\frac 1{4\pi r_0^3/3}=\frac{2k_F^3}{3\pi^2}.
\end{equation}

As the finite nuclei are in states near the standard one, the equation 
of state can be written approximately as\cite{Myers69}\cite{CWSZ00}
\begin{equation}
\label{EoSa}
 e(\rho_N,\delta)=-a_1+\frac 1{18}\big(K_0+K_s\delta^2\big)
 \Big(\frac{\rho_N-\rho_0}{\rho_0}\Big)^2+\Big[J+\frac L3
 \Big(\frac{\rho_N-\rho_0}{\rho_0}\Big)\Big]\delta^2,
\end{equation}
which is specified by the standard density $\rho_0$, volume energy 
$a_1$, symmetry energy $J$, incompressibility $K_0$, density symmetry 
$L$ and symmetry incompressibility $K_s$. These 6 quantities can be 
expressed in terms of 5 model parameters $C_\sigma^2$, $C_\omega^2$, 
$C_\rho^2$, $b$ and $c$\cite{CWSZ}. Reversebly, the 5 model parameters 
can be fixed if 5 of those quantities are known. In this case, we will 
choose $r_0$, $a_1$, $K_0$, $J$ and $K_s$ as input data, where $r_0$ is 
equivalent to $\rho_0$. Specifically, the procedure is as follows.

At the standard state $(\rho_0, 0)$, $e(\rho_0, 0)=-a_1$, and 
${\cal E}_\rho=0$, we have
\begin{equation}
{\cal E}_k+{\cal E}_\sigma+{\cal E}_\omega=\rho_0(M-a_1).\end{equation}
In addition, the equilibrium condition, $p(\rho_0, 0)=0$, can be written 
as
\begin{equation}
\frac 13({\cal E}_k-M\xi\rho_s)-{\cal E}_\sigma+{\cal E}_\omega=0.
\end{equation}
The following formulas can be derived from above two equations:
\begin{equation}\label{C2omega}
C_\omega^2=\frac{2M^2}{\rho_0^2}{\cal E}_\omega=\frac{M^2}{\rho_0^2}
\Big[\rho_0(M-a_1)-\frac 13(4{\cal E}_k-M\xi\rho_s)\Big],\end{equation}
\begin{equation}
{\cal E}_\sigma=\frac 12\Big[\rho_0(M-a_1)-\frac 13(2{\cal E}_k
+M\xi\rho_s)\Big].\end{equation}
Furthermore, Eqs. (\ref{Esigma}) and (\ref{Eqxi}) can be combined to 
solve the parameters $b$ and $c$ as
\begin{equation}
b=\frac{12{\cal E}_\sigma}{M^4(1-\xi)^3}-\frac{3\rho_s}{M^3(1-\xi)^2}
-\frac 3{C_\sigma^2(1-\xi)},\end{equation}
\begin{equation}
c=-\frac{12{\cal E}_\sigma}{M^4(1-\xi)^4}+\frac{4\rho_s}{M^3(1-\xi)^3}
+\frac 2{C_\sigma^2(1-\xi)^2}.\end{equation}
Finally, the equation for symmetry energy $J$\cite{CWSZ} is solved to 
give
\begin{equation}C_\rho^2=\frac{2M^2}{\rho_0}\Big(J-\frac 16
\frac{k_F^2}{\sqrt{k_F^2+M^2\xi^2}}\Big).\end{equation}

For given $r_0$, $a_1$, $\xi$, $C_\sigma^2$ and $J$, the above formulas 
can be used to calculate $C_\omega^2$, $C_\rho^2$, $b$ and $c$, and thus 
to calculate $K_0$, $K_s$ and $L$\cite{CWSZ}. According to the calculated 
$K_0$ and $K_s$, we can fix $\xi$ and $C_\sigma^2$. The following 
experimentally acceptable values\cite{CWSZ00} are used as input data in 
our calculation: 
\begin{equation}r_0\approx 1.14 fm,\,\,\,\,a_1\approx 16 MeV.
\end{equation} 

Fig.1 plots the calculated $K_s$ {\it versus} $K_0$ for given $\xi$. The 
solid curves from top to bottom correspond to $\xi=0.5$, $0.6$, $0.7$, 
$0.8$ and $0.85$, respectively. It can be seen that $K_s$ is negative 
only for $\xi$ larger than about $0.7-0.8$, if $K_0\le 300$ MeV is 
assumed. This is shown more clearly in Fig.2, where the calculated $K_s$ 
{\it versus} $\xi$ is displayed for given $K_0$. The solid curves from 
top to bottom correspond to $K_0=200$, $300$, $400$ and $500$ MeV, 
respectively. It is shown that $K_s$ is negative when $\xi$ is larger 
than about $0.73$, for $K_0\le 300$ MeV.

Fig.3 is the calculated $C_\sigma^2$ {\it versus} $K_0$ for given $\xi$. 
The solid curves from top to bottom correspond to $\xi=0.5$, $0.6$, 
$0.7$, $0.8$ and $0.85$, respectively. Fig.4 gives the calculated 
$C_\omega^2$ {\it versus} $\xi$. It can be seen from Eq.(\ref{C2omega}) 
that $C_\omega^2$ depends only on $\xi$, for given $r_0$ and $a_1$. The 
calculation of $C_\rho^2$ depends, beside $\xi$, also on the input value 
of $J$ and Fig.5 displays the calculated $C_\rho^2$ {\it versus} $\xi$ 
for given $r_0$, $a_1$ and $J=30$ MeV. On its turn, $L$ depends, beside 
$\xi$ and $J$, also on the input value of $K_0$, and Fig.6 shows the 
calculated $L$ {\it versus} $\xi$ for given $r_0$, $a_1$, $J=30$ MeV and 
$K_0$. The solid curves from top to bottom correspond to $K_0=200$, 
$300$, $400$ and $500$ MeV, respectively.

Fig.7 presents the nonlinear coefficient $b$ {\it versus} $K_0$ for 
given $\xi$. On the right hand side of the plot, the first three curves 
correspond to $\xi=0.5$, $0.6$ and $0.7$ from top to bottom, 
respectively. In the middle of the plot, the lower two curves correspond 
to $\xi=0.8$ and $0.85$ from top to bottom, respectively, being the 
first curve scaled by $\times 1/2$ and the second one by $\times 1/10$. 

Fig.8 displays the nonlinear coefficient $c$ {\it versus} $K_0$ for 
given $\xi$. On the right hand side of the plot, the solid curves 
correspond to $\xi=0.5$, $0.6$, $0.7$, $0.8$ and $0.85$ from bottom to 
top, respectively. The value of $c$ should be scaled by $\times 1/10$ 
for the curve of $\xi=0.8$, while by $\times 1/50$ for the curve of 
$\xi=0.85$. It can be seen from Fig.8 that $c$ is positive if $\xi$ is 
larger than about $0.7-0.8$, for $K_0\le 300$ MeV.

In addition to these results, it is worthwhile to see what could be 
obtained, if the realistic nuclear matter properties, extracted from 
measured data of finite nuclei by nonrelativistic models, are used as
input data. In this case, the results given by Myers-Swiatecki 
phenomenological nucleon-nucleon interaction\cite{Myers96}, Skyrme 
interaction\cite{Brack} as well as Tondeur interaction\cite{Tondeur} are 
employed.

The results of the calculation is presented in Table \ref{TableI}. The 
input data set $(r_0, a_1, K_0, J, K_s)$ is taken from the compilation 
of Ref.\cite{CWSZ01}. MS is for the Myers-Swiatecki interaction, SIII, 
Ska, SkM, SkM$^*$ and RATP for the Skyrme interaction, and Tondeur for 
the Tondeur interaction. It is worthwhile to note that the input value 
of $K_s$ is negative for all of these interactions. This is in agreement 
with the most expectations based on the nonrelativistic model\cite{Li}. 
As it can be seen from Eq.(\ref{EoSa}), physically, $K_s<0$ means that 
the compression modulus of nuclear matter will decrease if the 
asymmetry $\delta$ of nuclear matter increases. Experimentally, $K_s$ 
extracted from the isoscalar giant monopole resonance energy is between 
$-566\pm 1350$ to $34\pm 159$MeV\cite{Shlomo}. It is also interesting to 
note that the effective nucleon mass $M^*/M$ is around $0.89$, which 
reproduces nicely the value given by nonrelativistic models of nuclei. 
In addition, the nonlinear coefficient $c$ is positive, which means the 
field system is stable for all of these parameter sets.

In summary, the effective nucleon mass $M^*$ is discussed with relation 
to the symmetry incompressibility $K_s$ of nuclear matter, based on the 
model parameters fitted to nuclear matter properties, in the 
$\sigma$-$\omega$-$\rho$ model of the relativistic mean field theory 
with nonlinear $\sigma$-meson self-interaction. It is shown that $M^*$ 
is larger than $0.73M$, if $K_s$ is assumed to be negative and nuclear 
matter incompressibily $K_0$ is kept less than $300$ MeV. Furthermore, 
it is shown also that the field system is stable, as there is a lower 
limit for the $\sigma$-meson self-interaction energy in this parameter 
region. Our conclusion is: a stable result can be obtained by 
using an appropriate model parameter set altogether with a judicious 
selection of nuclear properties, within the $\sigma$-$\omega$-$\rho$ 
model of the relativistic mean field theory with nonlinear 
$\sigma$-meson self-interaction, at least for nuclear matter properties.

\begin{figure}
\caption{Calculated $K_s$ {\it versus} $K_0$ for given $r_0=1.14$ fm, 
$a_1=16$ MeV and $\xi$. The solid curves from top to bottom correspond 
to $\xi=0.5$, $0.6$, $0.7$, $0.8$ and $0.85$, respectively.}
\label{Figure1}
\end{figure}

\begin{figure}
\caption{Calculated $K_s$ {\it versus} $\xi$ for given $r_0=1.14$ fm, 
$a_1=16$ MeV and $K_0$. The solid curves from top to bottom correspond 
to $K_0=200$, $300$, $400$ and $500$ MeV, respectively.}
\label{Figure2}
\end{figure}

\begin{figure}
\caption{Calculated $C_\sigma^2$ {\it versus} $K_0$ for given $r_0=1.14$ 
fm, $a_1=16$ MeV and $\xi$. The solid curves from top to bottom 
correspond to $\xi=0.5$, $0.6$, $0.7$, $0.8$ and $0.85$, respectively.}
\label{Fifure3}
\end{figure}

\begin{figure}
\caption{Calculated $C_\omega^2$ {\it versus} $\xi$ for given $r_0=1.14$ 
fm and $a_1=16$ MeV.}
\label{Fifure4}
\end{figure}

\begin{figure}
\caption{Calculated $C_\rho^2$ {\it versus} $\xi$ for given $r_0=1.14$ 
fm, $a_1=16$ MeV and $J=30$ MeV.}
\label{Fifure5}
\end{figure}

\begin{figure}
\caption{Calculated $L$ {\it versus} $\xi$ for given $r_0=1.14$ fm, 
$a_1=16$ MeV, $J=30$ MeV and $K_0$. The solid curves from top to bottom 
correspond to $K_0=200$, $300$, $400$ and $500$ MeV, respectively.}
\label{Fifure6}
\end{figure}

\begin{figure}
\caption{Nonlinear coefficient $b$ {\it versus} $K_0$ for given 
$r_0=1.14$ fm, $a_1=16$ MeV and $\xi$. The solid curves correspond to 
$\xi=0.5$, $0.6$, $0.7$, $0.8$ and $0.85$, respectively.}
\label{Fifure7}
\end{figure}

\begin{figure}
\caption{Nonlinear coefficient $c$ {\it versus} $K_0$ for given 
$r_0=1.14$ fm, $a_1=16$ MeV and $\xi$. The solid curves correspond to 
$\xi=0.5$, $0.6$, $0.7$, $0.8$ and $0.85$, respectively.}
\label{Fifure8}
\end{figure}

\begin{table}[h]
\caption{The nuclear matter properties $r_0$(fm), $a_1$(MeV), $J$(MeV), 
$K_0$(MeV), $L$(MeV), and the parameters $C_\sigma^2$, $C_\omega^2$, 
$C_\rho^2$, $b$, $c$ of nonlinear $\sigma$-$\omega$-$\rho$ model in the 
relativistic mean field theory. See text for details.}
\begin{tabular}{lrrrrrrrrrrrr}
&$r_0$&$a_1$&$K_0$&$J$&$K_s$&$L$&$M^*/M$&$C_\sigma^2$&$C_\omega^2$&$C_\rho$&$b$&$c$\\
\hline
MS     &1.140&16.24&234.4&32.65&-147.1&85.553&0.893371& 92.728&30.908&27.729&-0.09203&1.1137\\
SIII   &1.180&15.86&355.5&28.16&-393.9&72.866&0.877399& 77.041&47.982&24.665&-0.15264&1.0935\\
Ska    &1.154&15.99&263.1&32.91&-78.45&86.766&0.885116& 96.522&38.415&29.434&-0.08115&0.8458\\
SkM    &1.142&15.77&216.6&30.75&-148.8&79.898&0.897339& 95.423&28.962&25.294&-0.08287&1.1499\\
SkM$^*$&1.142&15.77&216.6&30.03&-155.9&77.733&0.897526& 94.984&28.842&24.267&-0.08441&1.1655\\
RATP   &1.143&16.05&239.6&29.26&-191.3&75.430&0.893634& 89.460&31.269&23.183&-0.10318&1.1808\\
Tondeur&1.145&15.98&235.8&19.89&-39.78&47.603&0.886214&107.753&36.378& 9.705&-0.04911&0.6885\\
\end{tabular}
\label{TableI}
\end{table}

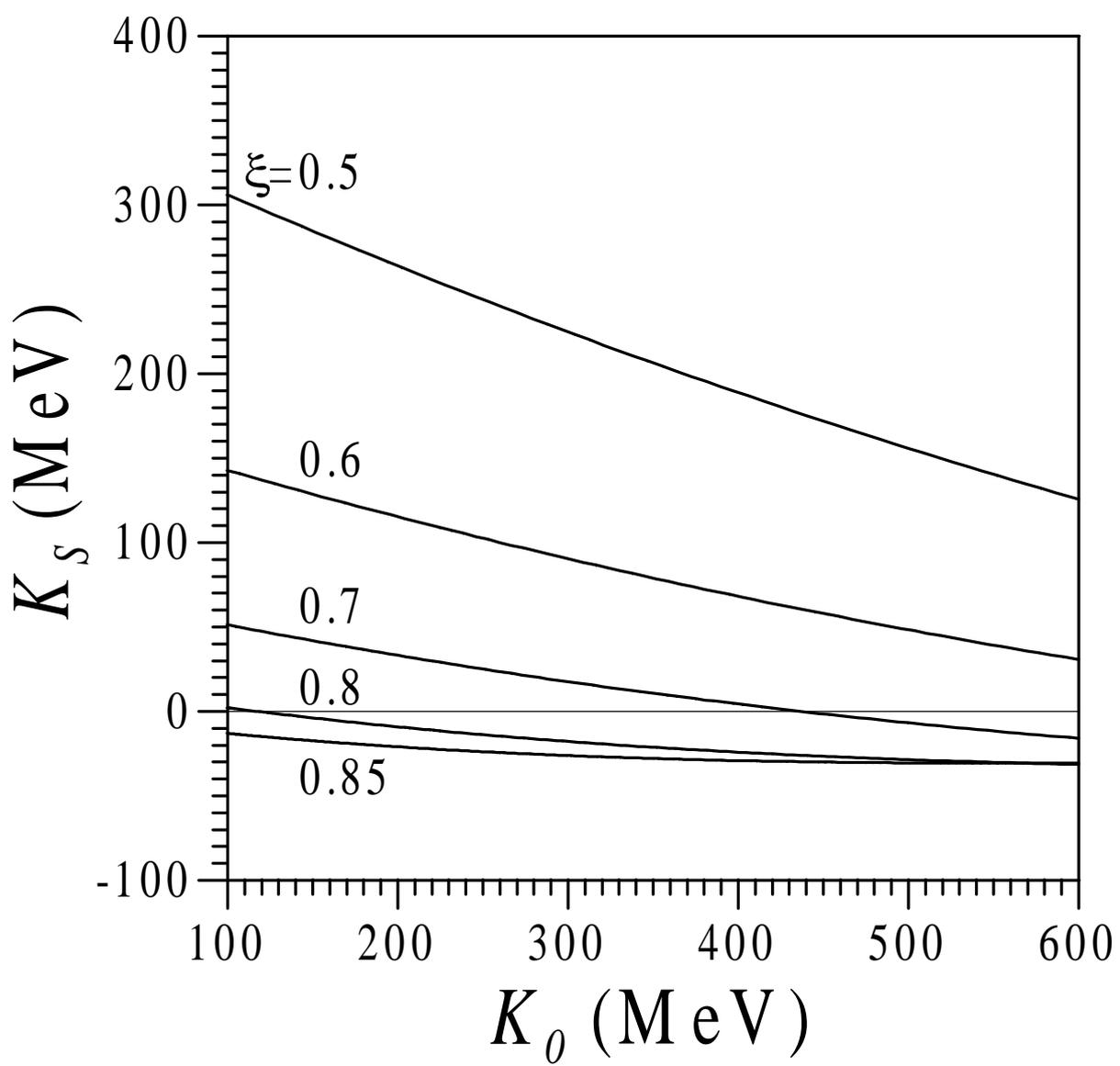
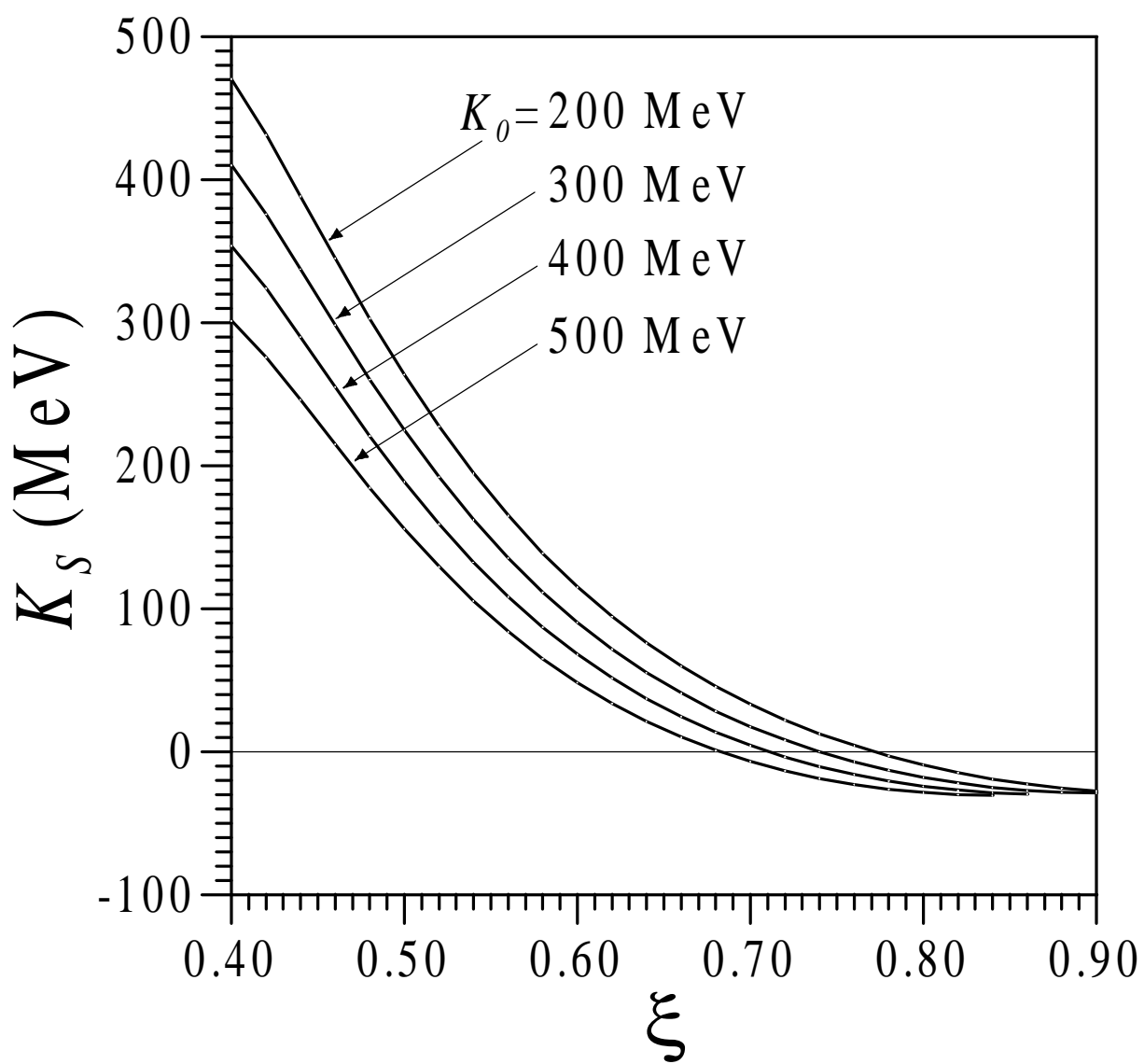
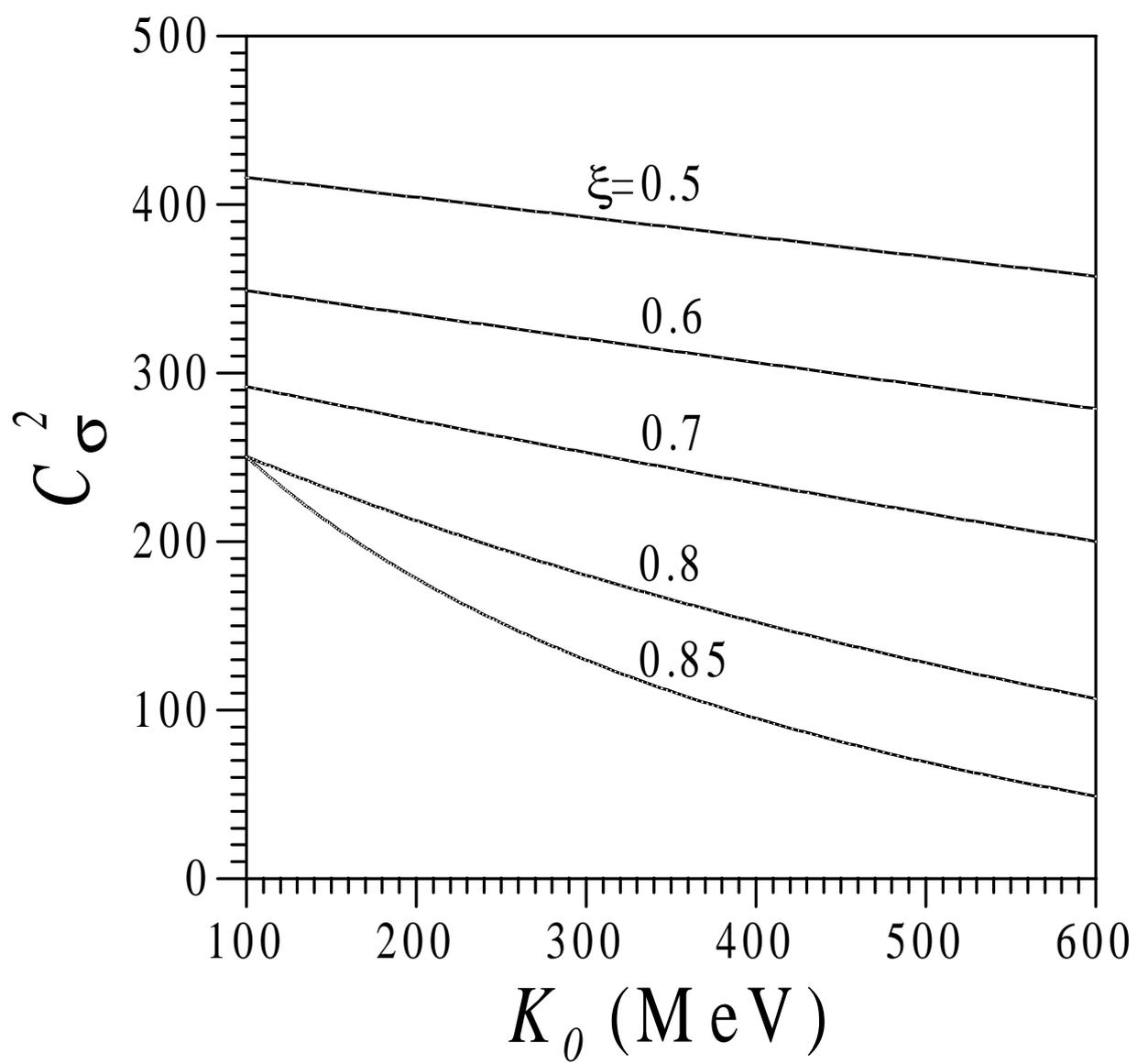
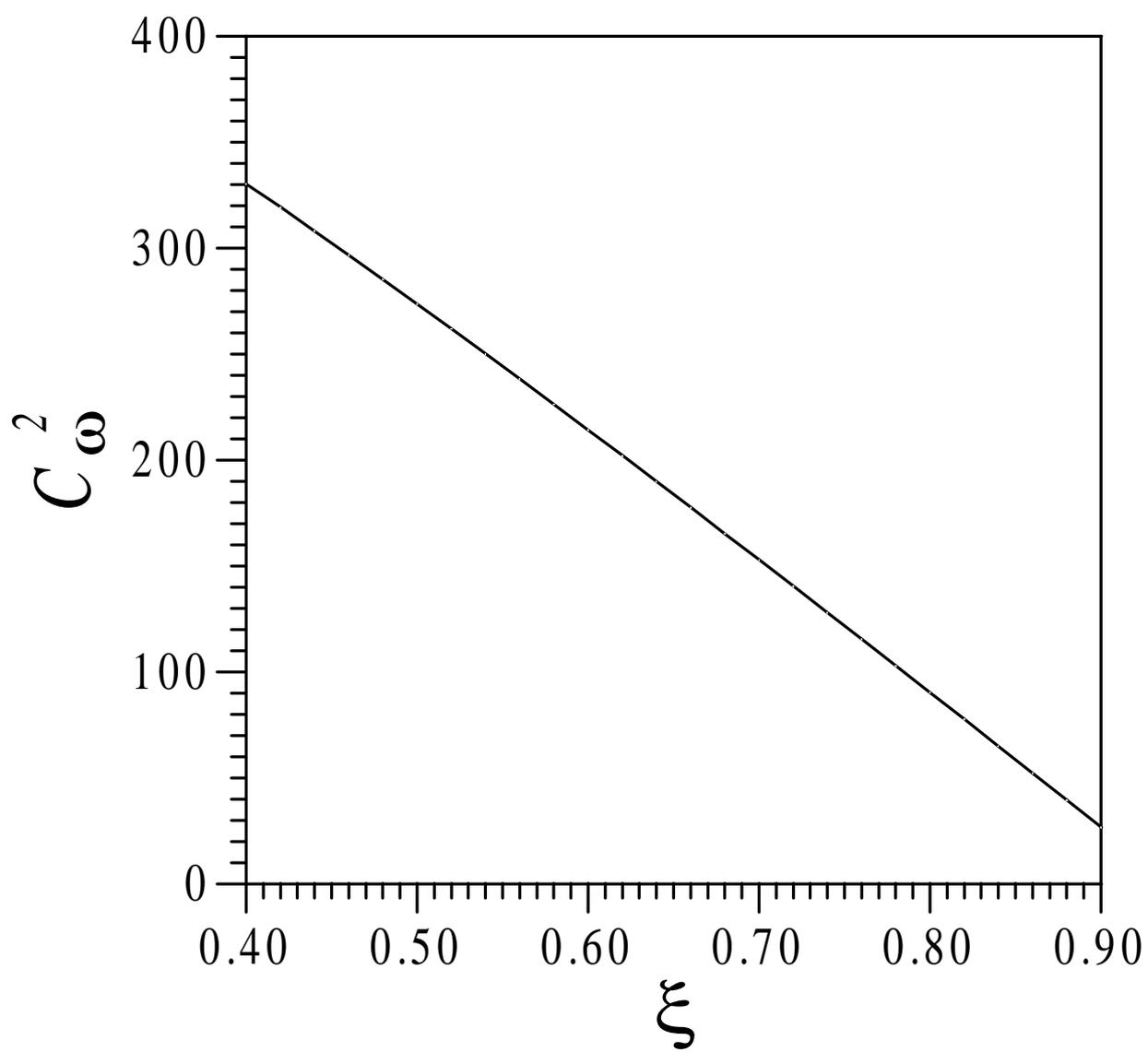
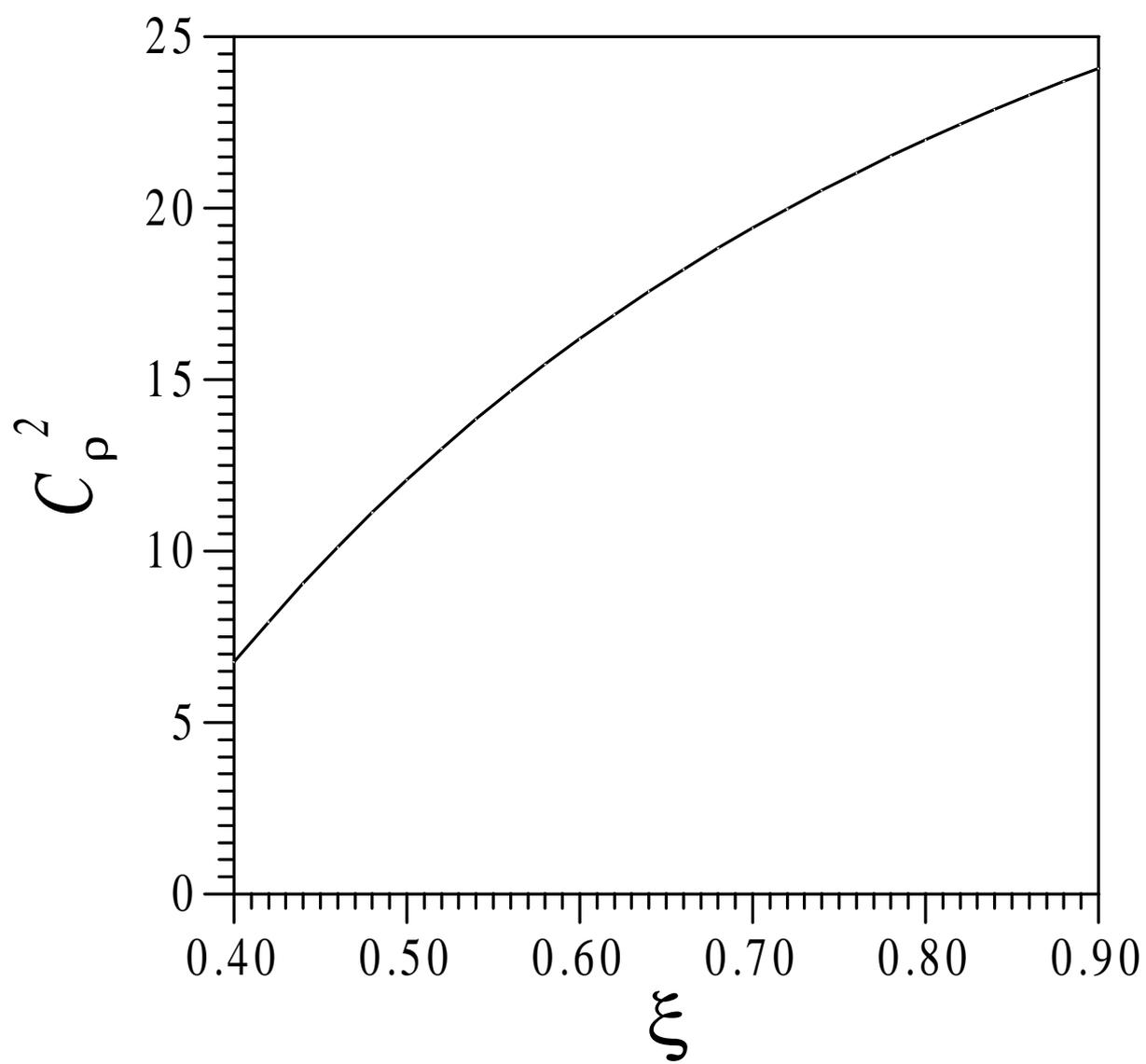
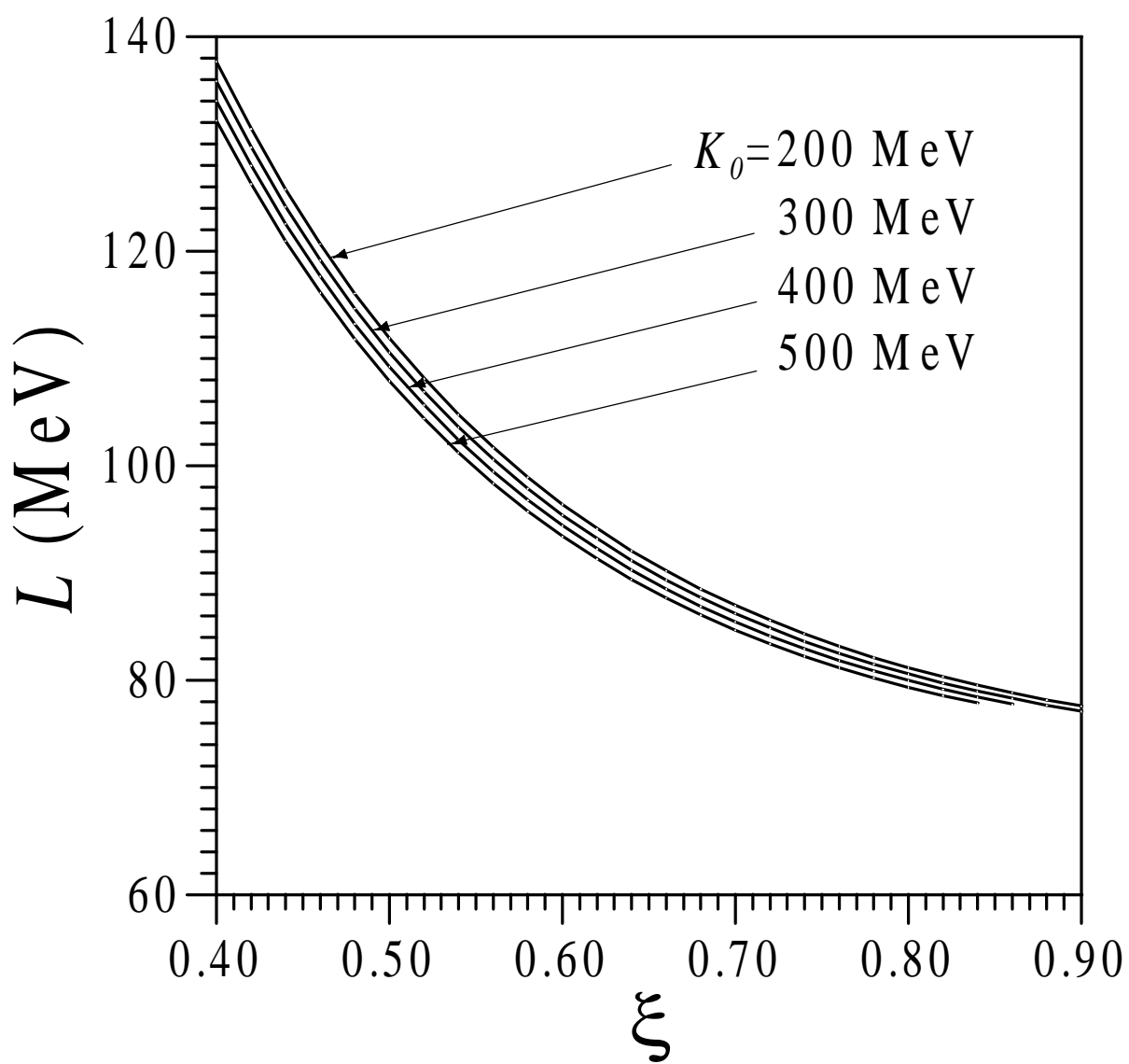
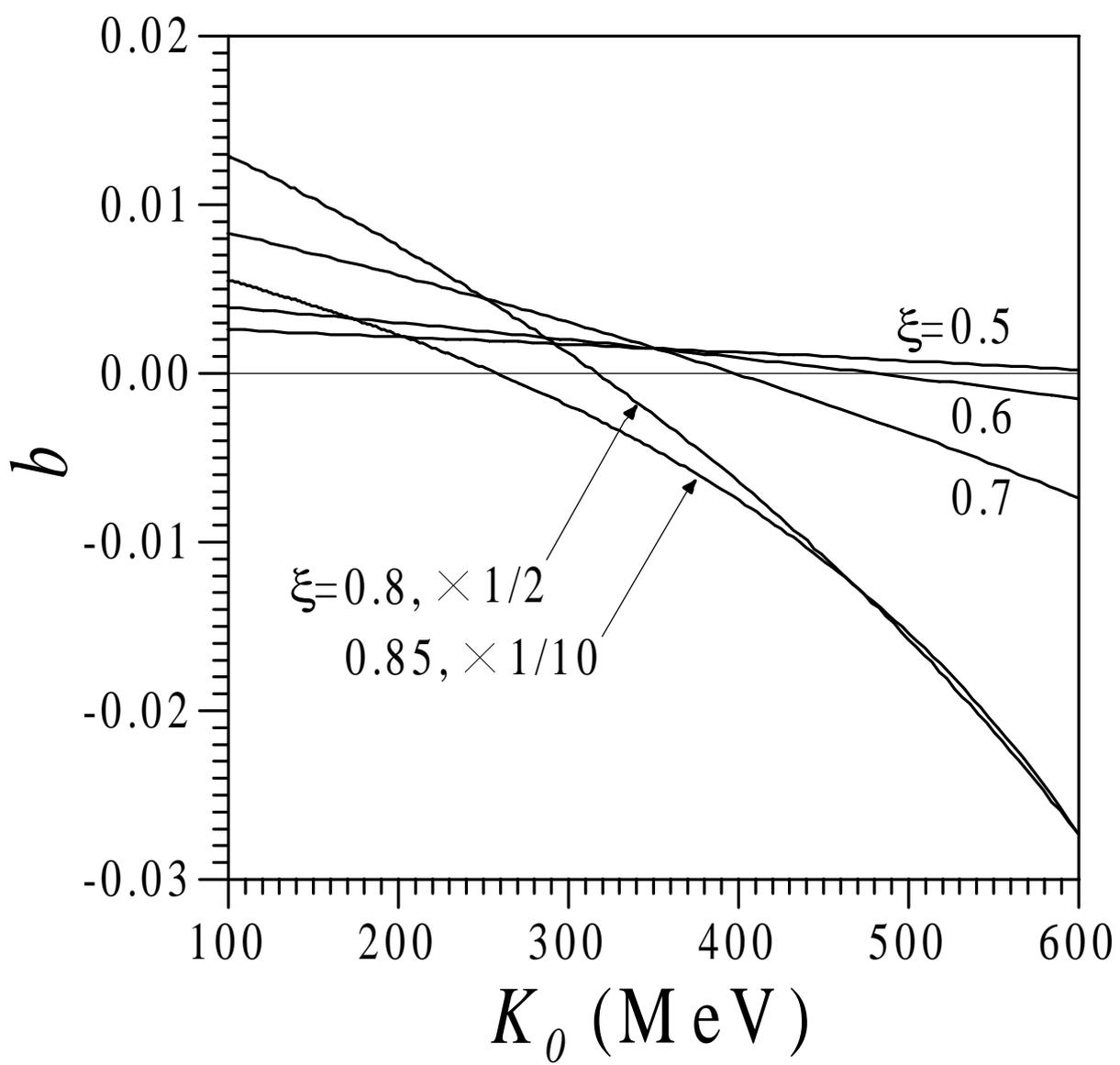
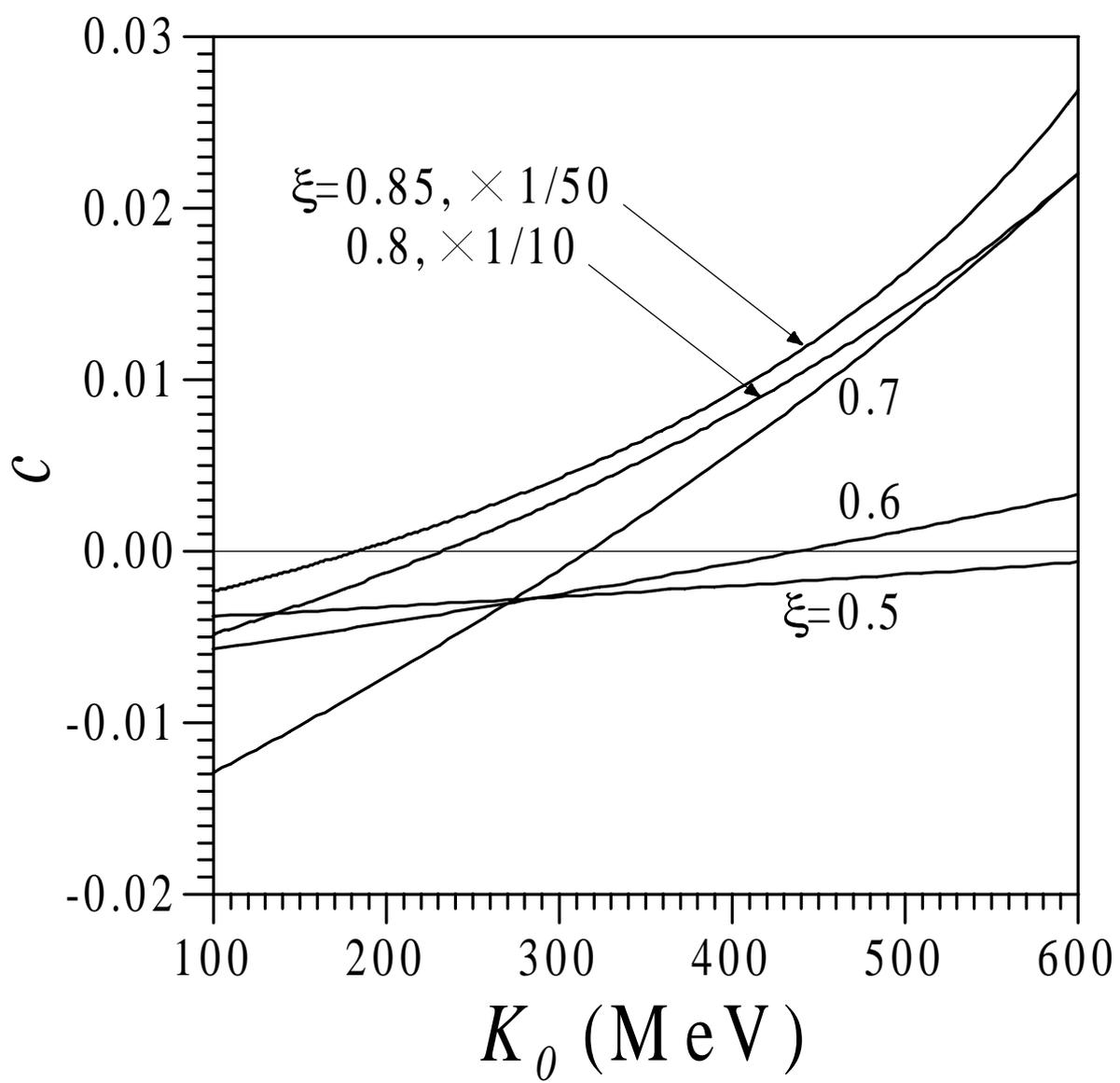
\end{document}